\documentclass[procedia]{easychair}

\usepackage{doc}
\usepackage{makeidx}
\usepackage{url}

\def\procediaConference{International Conference on Computational Science (ICCS 2016)}

\title{Teaching Data Science}

\titlerunning{Teaching Data Science}

\author{
    Robert J. Brunner\inst{1}
\and
    Edward J. Kim\inst{2}
}

\institute{
  University of Illinois,
  Urbana, Illinois, U.S.A.\\
  \email{bigdog@illinois.edu}
\and
   University of Illinois,
   Urbana, Illinois, U.S.A.\\
   \email{jkim575@illinois.edu}\\
 }

\authorrunning{Brunner and Kim}

\begin{document}

\maketitle


\begin{abstract}
    We describe an introductory data science course, entitled
\textit{Introduction to Data Science}, offered
at the University of Illinois at Urbana-Champaign. 
The course introduced general programming concepts by using the Python programming language
with an emphasis on data preparation, processing, and presentation.
The course had no prerequisites, and students were not expected to have
any programming experience.
This introductory course was designed to cover a wide range of topics, from the nature of
data, to storage, to visualization, to probability and statistical analysis,
to cloud and high performance computing,
without becoming overly focused on any one subject. We conclude this article with a discussion of lessons learned and our plans to develop new data science courses.
\end{abstract}


%
%

\section{Introduction}
\label{sec:introduction}

Data scientist has been dubbed ``the sexiest job of the 21st
century"~\cite{davenport2012data}.
A report from McKinsey Global Institute estimates that by 2018
the United States alone could experience a shortage of 190,000 data scientists
with deep analytical skills, and 1.5 million data-savvy managers
and analysts capable of taking full advantage of Big Data~\cite{mckinsey}.
Whether or not such media coverage is hype, a surge in the popularity of
data science will naturally create a demand from students for training in 
data science.
Thus, it is not surprising that numerous data science programs have emerged at
major universities, including the University of California at Berkeley, Stanford, New York University, and the University
of Michigan.

At the University of Illinois at Urbana-Champaign, we began offering an
eight-week online course,
\textit{Informatics 490: Introduction to Data Science}, in the Summer of 2014.
This new course was not our initial foray in teaching data science concepts.
In 2010, we began adding data science concepts into an undergraduate astronomy
course, \textit{Astronomy 406: Galaxy and the Universe}. Subsequently, with the assistance of the Illinois Informatics
Initiative, we created a new course in the Fall of 2011,
\textit{Astronomy 496/596: Practical Informatics for the Physical Sciences}
\cite{brunner2013practical}.
Although these previous efforts were generally successful, their target audience
were students in science and engineering disciplines. Given their popularity, the University of
Illinois community recognized the need for this material to be offered more broadly.
As a result, we created and taught \textit{Introduction to Data Science} as an online course in the Summer of 2014. 
This new course was successful, and was expanded and offered again in the Spring of 2015, but 
as a fifteen-week, dual-level undergraduate/graduate course.

The new, semester-long course differed from typical courses offered in statistics
or computer science departments. Our new course taught the Python programming language
with more emphasis on data preparation, processing, and presentation.
Many applied statisticians, including Nolan and Temple Lang~\cite{nolan2012computing},
Cobb~\cite{cobb2015mere}, and Donoho~\cite{donoho2015fifty}, lament that
computing concepts are not a major component of traditional statistics.
Furthermore, a modern, practical data science course offered within
statistics curricula will most likely use the R language~\cite{hardin2014data}.
Python (especially when using the Pandas library) is capable of performing most, if not all,
of the data analysis operations that a data scientist might complete by using R. 

Python, however, is a general
purpose programming language,
while R is generally limited to tasks within the realm of statistical computing.
With Python, students learn and build confidence using general programming concepts that are common amongst other popular programming languages such as Java, C, or C++.
The course also differs from typical introductory computer science courses
that emphasize program design, as we integrate general programming concepts
with an emphasis on data preparation, processing, and presentation.
Moreover, data-intensive computer science courses, such as a machine learning
course, often have one or more programming classes as a prerequisite, which becomes a barrier
to enrollment for students in some fields, such as the social sciences.
Our course has no prerequisites, and students are not expected to have any programming experience.

In the end, fifty students completed the three-credit course in the Spring
of 2015.
Students were from diverse disciplines: Statistics, Physics,
Computer Science,  Economics, Electrical Engineering,
Technical Systems Management, Political Science, Industrial Engineering,
Mathematics, Nuclear Engineering, Astronomy, Actuarial Science,
Library and Information Science, Material Science,
Advertising, Agricultural Engineering, Biological Engineering, Biophysics,
Community Health, Finance, Food Science, Geographic Information Sciences,
Graphic Design, Linguistics, Material Science, Psychology, and Sociology.
With students from such a diversity of backgrounds, we feel that the course clearly filled
a pressing need.

In the rest of this paper, we detail our experiences in teaching this fifteen-week
data science course. After which we discuss some general lessons learned as well as our plans for
the future.
Course materials are available at \url{https://github.com/UI-DataScience/info490-sp15}.

\section{Course Structure}
\label{sec:course_structure}

Our course was designed to cover a wide range of topics, from the nature of
data, to storage, to visualization, to probability and statistical analysis,
to cloud and high performance computing,
without diving into any one subject too deeply.
This was done to ensure that students would be exposed to as many new topics
as possible with sufficient exposure so that they could independently
explore topics of interest as desired.

Overall, the class was built by using a Moodle installation operated by the
College of Liberal Arts and Sciences at the University of Illinois.
This provided the students with a common location for obtaining all lecture
material, supplementary material (including readings and code examples),
and online forums to facilitate class communication.
However, we also forced the students to use the \texttt{git} versioning tool by spending
an entire lecture on Git and by hosting Jupyter notebooks for lectures and
assignments on GitHub.
Since this course was delivered entirely online via Moodle and GitHub, our course resembled
a massive online open course (MOOC) in several aspects.

From the student's perspective, each week began with an introduction video in which we summarized the topics
that would be covered, and dynamically demonstrated example codes or applications.
Each week's material was organized into three lessons; each lesson was
comprised of a required reading list and one or more optional readings.
For some of the lessons, we provided Jupyter notebooks~\cite{jupyter}
(formerly IPython~\cite{perez2007ipython}; see Section~\ref{sec:python}),
while in others we leveraged freely available, online resources.
At the end of each lesson, there was a short lesson assessment with
ten multiple-choice questions, and at the end of each week, there was a comprehensive quiz with forty
multiple-choice questions. In addition, there was a programming assignment for weeks two through fourteen (this was a fifteen week course), where Jupyter
notebook templates were provided to assist students in preparing their solutions.
All programming assignments were graded via peer assessment and where possible via automated machine grading (see Section~\ref{sec:assignment}).
Given the online nature of the course, there was no midterm or final exam.

\section{Course Content}
\label{sec:course_content}

In this section, we detail the topics covered in the course within their own
subsections.
The presentation order roughly matches the order the material was presented
in the course.

\subsection{UNIX and the Command Line Interface}
\label{sec:unix_cli}

One of the major benefits of using Python is the large number of libraries that
are available; however, this can present a stumbling block for students who are
uncertain on how to acquire, build, and install new code, libraries, or applications.
In a conventional classroom setting, a practical course such as ours would
employ a computer laboratory, where all necessary software would have been installed.
However, since we created an online course, students had
to use their own machines. Thus, we needed a solution that removed the
difficulties associated with installing and upgrading software packages.
Although the Anaconda Python distribution was one possible solution,
we instead chose to use the Docker container technology~\cite{docker}, because
our course would start with an introduction to UNIX and
the Command Line Interface (CLI).

Data scientists are expected to be at least familiar with a CLI because
sometimes shell commands are simply the best tool for cleaning and
manipulating various types of data.
In an announcement we sent out to enrolled students one week prior to
the beginning of the semester, students were instructed
to follow the provided documentation to install the Docker technology, which depended on their host computer operating system.
We also built and provided a Docker course image that contained all software
and libraries needed for the course. Given the importance of Docker in our course, the very first reading assignment
introduced the concept of virtualization and basic Docker usage.

In the following two weeks, students read sections from the
\textit{The Linux Command Line}~\cite{shotts2012linux} book
by entering commands into a terminal window within their Docker container.
After two weeks, students had a basic understanding of the UNIX file system,
understood how UNIX processes work, learned how to use \texttt{vim} to
perform basic file editing, and were able to use data processing tools such as,
\texttt{grep}, \texttt{awk}, and \texttt{sed}.
Since we hosted Jupyter notebooks for lessons and programming assignments on
GitHub, we also provided reading assignments on versioning and \texttt{git}.

\subsection{The Python Programming Language}
\label{sec:python}

Since our course had no prerequisites, we devoted two weeks to the general
concepts of programming.
For this, students were asked to read from
\textit{Think Python} \cite{downey2012think}. This introduced conditionals, functions, the basic Python data types,
the Python object oriented approach to data, and Python data structures such as the tuple, list, and dictionary.

\begin{figure}
	\begin{centering}
	\includegraphics[width=0.9\textwidth]{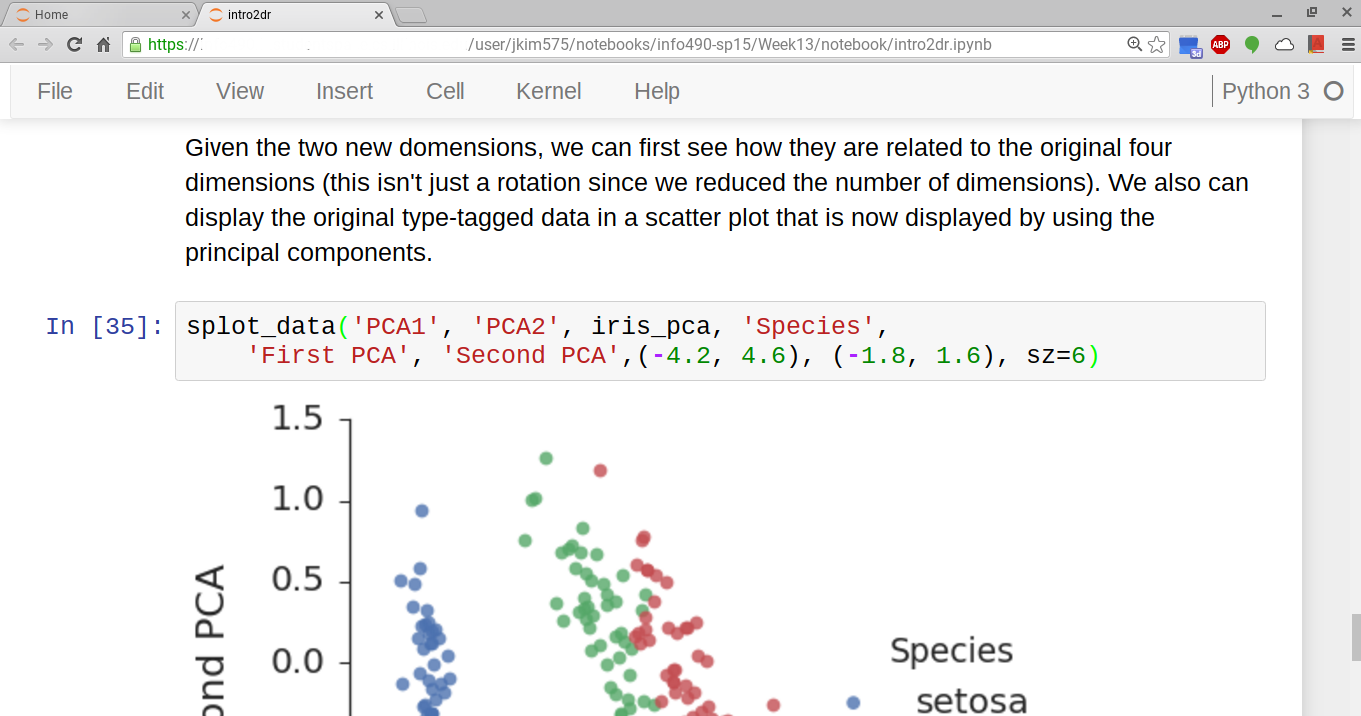}
	\caption{An example session of Jupyter notebook.}
	\label{fig:jupyter}
	\end{centering}
\end{figure}

Many reading assignments were provided to students as Jupyter
notebooks~\cite{jupyter}, and students completed all Python programming
assignments by using Jupyter notebooks.
The Jupyter notebook is an interactive computational environment,
in which you can present and execute code, include descriptive text, and inline visualizations in a
single document as shown in Figure~\ref{fig:jupyter}. This technology is becoming popular as an instructional tool, and although its interface
sometimes can complicate debugging, we have found it to be an excellent tool for teaching data
science.

\subsection{Data Visualization}
\label{sec:visualization}

The next phase of the course introduced one of the most important yet least
discussed aspects of data science, namely data visualization.
Although only one week was explicitly devoted to visualization, we introduced
additional visualization ideas and examples throughout the 
rest of the course wherever appropriate.

To demonstrate the importance of plotting data before analyzing it,
we showed Anscombe's quartet~\cite{anscombe1973graphs}.
We also presented ideas championed by Edward Tufte, such as
minimizing chart junk---that is, just because a software package can add
unnecessary embellishments to a plot does not mean the user should do so.
We also linked to a video of McCandless's 2010 TED talk,
\textit{The beauty of data visualization}, which demonstrates how
hidden patterns can emerge from beautiful, simple visualizations of complex data
sets.

Visualizing data with Python required the introduction of the
Numpy~\cite{walt2011numpy} and Matplotlib~\cite{hunter2007} libraries,
which we used to demonstrated how to plot a theoretical curve,
how to make a scatter plot from a data set, and how to make a histogram.
We demonstrated how to use the Seaborn~\cite{seaborn} library to improve the appearance
of a plot, and, throughout the rest of the course,
we used Seaborn to create a variety of plots, such as
box plots, violin plots, kernel density estimation plots, pair plots, distribution
plots, time series plots, and heat maps.

\subsection{Statistical and Time Series Analysis}

We devoted three weeks to statistical and time series analysis.
First, we introduced the Pandas library, which introduces concepts from the
R language into Python to greatly simplify working with large, structured
data sets.
The primary object in Pandas is the DataFrame, a two-dimensional
tabular data structure with both row and column labels. The DataFrame 
provides sophisticated functionality to make it easy to select, group,
and filter data. 

After an introduction to Pandas, we transitioned to statistical analysis,
which is very important with large data sets, especially when we need to make a
quick decision based on summary information without being able to completely
scan a (potentially large) data file.
It is also necessary to use statistical analysis when we wish to model some
data component in order to make future predictions or to better understand
some physical process.
All of these cases require an understanding of summary statistics,
density functions, distributions functions, basic Bayesian analysis,
and basic statistical analysis including linear modeling and regression.

For most of this section of the course, we leveraged the freely available
books, \textit{Think Stats}~\cite{downey2014think}
and \textit{Think Bayes}~\cite{downey2013think}, by Allen Downey.
While a considerable code base accompanies these books, we encouraged
students to not use the author's code, but to instead use standard Python, Numpy,
or Pandas data structures.
This was because Downey's code is written like a software project and is
spread across multiple files and classes.
While his approach makes the code quite robust and easy to extend, we felt it complicated learning statistical concepts, especially for students who have little previous programming experience.

Finally, we devoted one week to learn how to work with time series data in Pandas
and how to visualize them by using Seaborn. 
Time series data are everywhere, and while often simple in content 
(e.g., a one-dimensional array indexed by time)
they often provide rich data mining opportunities. 
As an example, the last lesson in this week focused on mining twitter streams
by using Twitter's API.
We also introduced Markov Chain Monte Carlo (MCMC) methods by leveraging
Chapter 1 of \textit{Bayesian Methods for Hackers}~\cite{davidson-pilon2015bayesian},
which demonstrates the use of PyMC~\cite{patil2010pymc} to sample from the
posterior to estimate model parameters.

\subsection{Data Formats}

\begin{figure}
	\begin{centering}
	\includegraphics[width=0.75\textwidth]{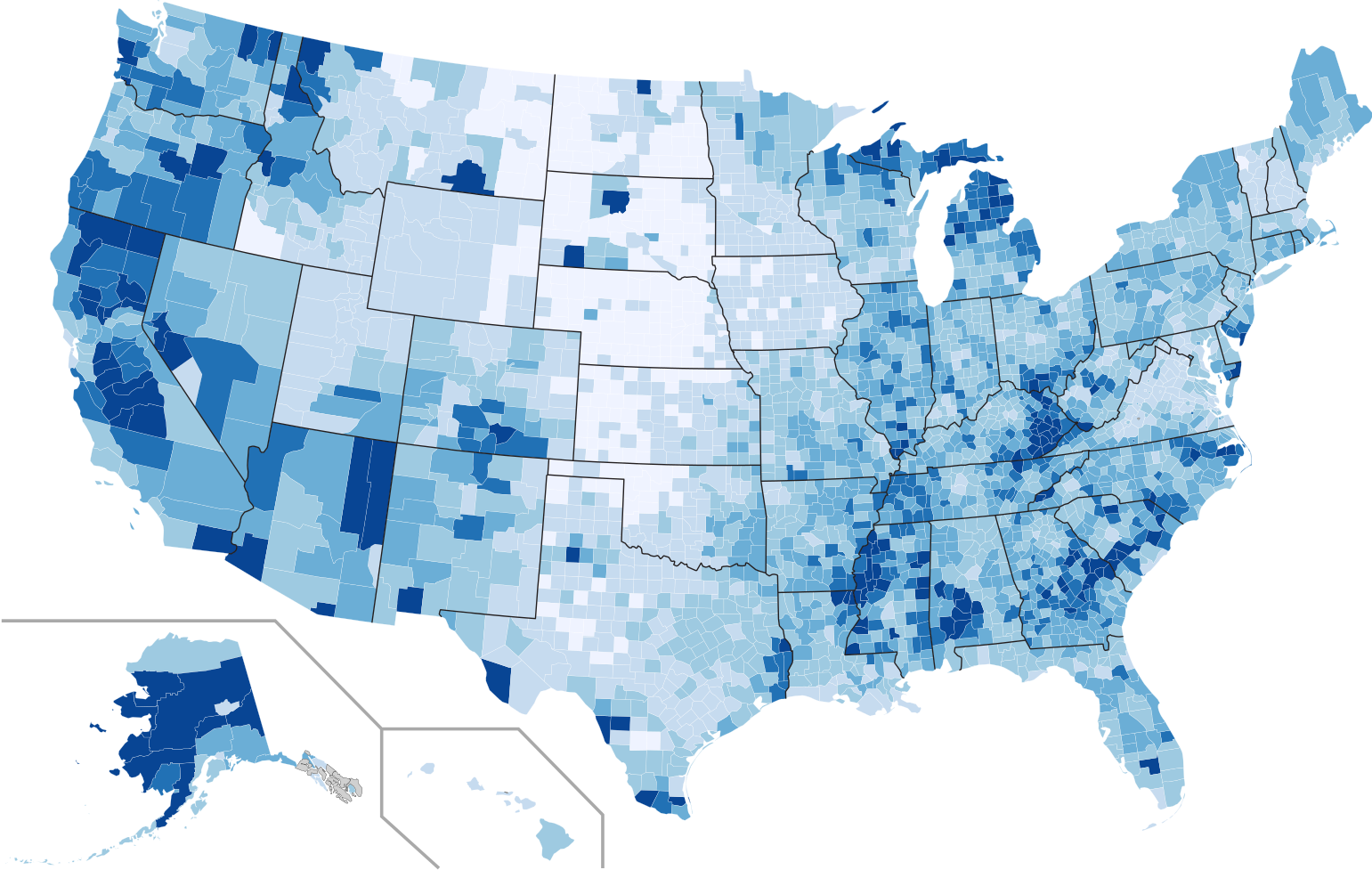}
	\caption{A Choropleth map that visualizes county-level unemployment rate in 2013.}
	\label{fig:choropleth}
	\end{centering}
\end{figure}

Recognizing and understanding how to use standard data formats is essential to being a data scientist.
Therefore, we devoted one week to basic data processing tasks and different data file
formats, such as fixed-width text, delimited text, XML, JSON, and HDF.
As a practical example, we demonstrated how to make a Choropleth visualization,
which was inspired by Nathan Yau~\cite{yau}.
To create a Choropleth map such as the one in Figure~\ref{fig:choropleth},
students needed to access web resources,
parse an XML-based data format, extract meaningful data from a second
web-accessible resource, and combine it all into a new and interesting
visualization.

\subsection{Data Persistence}

Relational database management systems (RDBMS) remain the most popular data
storage technology. Thus, they are essential to any data science course, and we devoted two
weeks to introduce this technology and to explore how to programmatically work with them.
First, we demonstrated how to perform simple data persistence via the Pickle
library.
This provided a simple introduction to the basic idea of storing data before
we introduced RDBMS.
Next, we introduced SQL and the SQLite database, by presenting basic SQL data types,
including the SQL NULL value, the SQL Data Definition Language (DDL),
the SQL Data Manipulation Language (DML), and the overall concept of a database schema.

Following one week of learning to work with SQL by using the SQLite database from a UNIX command line, we transitioned to using
Python and SQLite to programmatically interact with a database.
We demonstrated how to establish a database connection from within a Python
program and how to use this connection to obtain a reference to a database
cursor.
Students learned how to dynamically interact with an SQLite database in
Python by issuing queries, and also by updating and inserting rows on
existing database.
Students also learned how they can use a Pandas DataFrame to simplify 
most of these tasks.
Finally, we provided reading assignments on NoSQL databases, including
the MongoDB and Apache Cassandra technologies.

\subsection{Introduction to Machine Learning}

Although machine learning is an enormous field that can easily fill an entire
semester course,
the simple, consistent API of the scikit-learn library~\cite{scikit-learn}
and its excellent documentation allowed us to devote one week to
an introduction of supervised learning methods
(e.g., $k$-Nearest Neighbors, Support Vector Machine, and Decision Trees),
dimensionality reduction (Principal Component Analysis), and
cluster finding ($k$-means).
While it is debatable whether machine learning belongs in a truly
introductory course such as ours, most students found the topic fascinating,
and we felt that by introducing machine learning, we could motivate students to independently
pursue further studies in advanced machine learning or to enroll in
existing offerings in computer science department.

\subsection{Introduction to HPC and Cloud Computing}

The final week of the course introduced computational technologies that
are often required in data intensive analyses.
We began by introducing the concept of cloud computing and instructed
students to register for a free trial of Google Compute Engine.
After reviewing the Docker engine, students learned how to run containers
of our course Docker image on the Google Compute Engine, which enabled
web-accessible Jupyter notebook servers in the cloud.
Finally, we demonstrated how to run a Hadoop Streaming MapReduce job
in a Docker container as well as the use of the Hadoop file system (HDFS)
and the Hadoop Streaming process model.

\section{Assignments}
\label{sec:assignment}

As mentioned previously, our course was delivered entirely online.
As a result, it was similar to a MOOC platform in various aspects.
Email and online forums were the primary form of communication
between students and instructors;
all assignments were submitted online;
programming assignments were graded by peers for sixty percent of the assignment
grade;
and for the remaining forty percent they were graded automatically,
except for some problems that were not amenable to unit testing.

There were three types of assignments: multiple-choice quizzes;
weekly programming assignments; and peer assessments.
Since a programming class has not been a prerequisite to this class, we
provided template code for each programming assignment.
During this process, we first wrote a Jupyter notebook that demonstrated
a data science task by using unit-testable functions. Next, we removed a few lines of code from one or more functions.
We provided a detailed description for each function, the data types of any input parameters and return values, and we provided several examples
of correct output given a specific input.
Except for a few visualization problems, all programming assignments were
graded automatically by using unit tests that compared student outputs against the
correct answers.

For peer assessments, at least four students were assigned to evaluate each
student's assignment.
We provided a grading rubric that detailed how points should be allocated
according to different aspects of the assignment, such as correctness and
readability.
We believe that there are significant benefits to peer assessments, as
students can learn from seeing various approaches to the same problem.
Furthermore, by autograding programming assignments and utilizing peer
assessment, the course could easily be scaled up, allowing even more students to enroll in future versions of this course.

In many programming assignments, we used the airline on-time performance data
set from the American Statistical Association (ASA) Data Expo
2009~\cite{asa,wickham2011asa}.
The data, which originally comes from the Bureau of Transportation
Statistics, tracks all commercial flights in the U.S.\ between 1987 and 2012.
The size of the data set, with more than 120 million records in total and
taking up twelve Gigabytes, precludes traditional, in-memory processing.
Therefore, we used the CSV file for the most notable year, 2001, which
contains around six millions rows and twenty-nine columns, and takes up 573 Megabytes.  

\begin{figure}
    \centering
    \begin{minipage}[b]{0.48\linewidth}
        \centering
	\includegraphics[width=\linewidth]{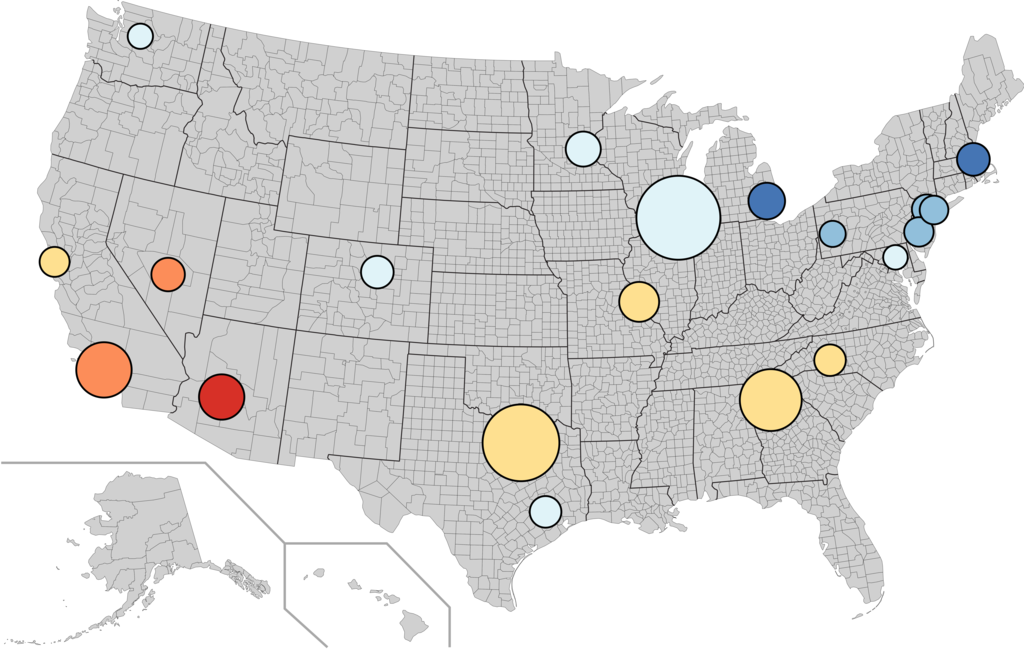}
	\caption{Top twenty U.S.\ airports. The size of each circle is proportional
            to the total number of arrivals and departures in 2001. The redder
            the color, the higher the temperature at the airport; the bluer the
            color, the lower the temperature.} 
	\label{fig:top20}
    \end{minipage}
    \hfill
    \begin{minipage}[b]{0.48\linewidth}
        \centering
        \includegraphics[width=\linewidth]{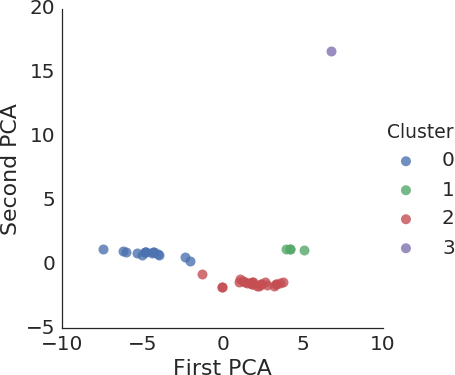}
        \caption{Clustering of Delta Airline's aircraft.}
        \label{fig:clustering}
    \end{minipage}
\end{figure}

One programming assignment asked students to use the 2001 data
to visualize the total number of flights and current temperature of the top
twenty airports (by traffic) in the U.S.\ as shown in Figure~\ref{fig:top20}.
Students read two columns, origin and destination airports, from this CSV file,
and used \texttt{groupby} and \texttt{aggregate} operations to aggregate the
count of flights at each airport.
Each airport in the original CSV file was identified by an IATA code
(a three-letter code designating many airports around the world). Thus, students had to map 
an airport in the SVG template file, which was identified by its city name,
e.g., Chicago or San Francisco, to an IATA code and then programmatically retrieve the current temperature for that city. 
This was completed by issuing an HTTP request and processing the XML response 
from the Federal Aviation Administration (FAA) web service. 
Students parsed the SVG file and searched for all instances of
\texttt{circle} in order to construct the target visualization:

\begin{verbatim}
<circle cx="367.68198" cy="118.095" stroke="black" stroke-width="1" r="5" fill="red"
id="Chicago"/>
\end{verbatim}

\noindent
Finally, students replaced the \texttt{r} variable with a number that was proportional
to the counts of flights at the corresponding airport, and set the \texttt{fill}
variable with a color to indicate the current temperature.

Another programming assignment asked students to perform Principal Component
Analysis (PCA) before applying the $k$-means algorithm to group similar
aircrafts into clusters as shown in Figure~\ref{fig:clustering}.
Delta Airlines (and other major airlines) host data on all of their aircraft on
their website\footnote{\url{
http://www.delta.com/content/www/en_US/traveling-with-us/airports-and-aircraft/Aircraft.html}}.
To provide students with data for the clustering task, we scraped the webpage and provided a CSV file with thirty-four features from forty-four different aircraft. 
These features included quantitative measurements such as cruising speed,
accommodation, range in miles, as well as categorical data such as whether a particular aircraft has Wi-Fi or video.

\section{Discussion}
\label{sec:discussion}

Developing and teaching this course was both a challenge and a rewarding experience. 
The primary challenge resulted from the need to find the right balance for students with diverse
programming abilities and backgrounds.
Some students had programming experience, some were already established Python programmers,
while others came in with no prior exposure to programming.
We tried to make the programming assignments not too burdensome for novice
programmers, but it was difficult to keep more advanced students
motivated. We also struggled to maintain a consistent level of difficulty and
workload when assembling course material from a variety of sources.

The course presented a lot of material in a very short time.
Although students who were willing to invest the necessary time and effort
indicated overwhelmingly in a post-class survey that they learned a great deal,
the heavy workload could turn away students who are new to programming
or those who are balancing heavy course loads. In addition, we found that one semester was inadequate to introduce all topics, especially applications of topics we did cover such as text analysis, network or graph analysis, or analysis of social media. 
Therefore, we decided to expand this existing course into two new courses:
\textit{Foundations of Data Science} which was first taught in the Fall 2015 semester, and
\textit{Advanced Data Science} which will be taught in the Spring 2016 semester. The material for both of these courses will be publicly available at \url{https://github.com/ui-datascience}, and we will once again leverage Docker technology and Jupyter notebooks. Enrollment for both of these courses has continued to grow, indicating the strong need for more courses.

By splitting the original course into two courses, we will be able to explore programming concepts at a slower pace 
in the first semester, and move on to more advanced topics,
such as machine learning and cloud computing, in the second semester course.
While the syllabus remains in flux, we currently plan to devote a significant fraction of the second semester
(five to six weeks) to machine learning, before moving on to a variety of advanced topics
such as text mining the social media, network analysis, probabilistic
programming, and cloud computing.

The Docker container technology has been a crucial tool that allowed us to
introduce the UNIX CLI and to ensure that every student
was working in an identical environment.
Although there are other virtualization tools, such as VirtualBox, we believe
Docker is a better solution given its lighter footprint and connection to modern cloud computing infrastructures.
However, running Docker on either Mac OSX or Windows requires another layer of virtualization. For example, 
Windows users have to use Boot2docker~\cite{boot2docker} or the
Docker Machine~\cite{docker-machine}.
These Windows tools are not yet mature, and Windows users faced various
problems during installation and at runtime (we note that Mac OSX users had fewer difficulties in this area, but still had more challenges than those who used a modern Linux distribution).

Since over forty percent of the students in this class were Windows users, and since
most students did not have the technical proficiency to troubleshoot software problems as they arose, we are now exploring using a cloud computing system to launch a server running
JupyterHub\footnote{https://github.com/jupyter/jupyterhub} for future course instances.
JupyterHub is a multi-user version of the Jupyter notebook on a centralized
server.
No installation is necessary on the part of the users; after they log into
the JupyterHub server, they have immediate access to their Jupyter notebooks
running inside a Docker container. This approach is easier on the end user and also provides the instructional staff with new opportunities, such as automated assignment collection and simpler updates to student Docker containers.
However, running a centralized server has the disadvantage that the instructional staff must secure sufficient computing resources, which can be challenging, especially for large courses.

In conclusion, we learned a great deal from developing and teaching this data
science course.
We hope that others can make use of our experiences and the course material we are now publishing to more broadly improve
the training of the next generation of data scientists.


\section{Acknowledgments}

We acknowledge support from the Illinois Informatics Initiative, the University of Illinois, and the National Science Foundation Grant No.\ AST-1313415. RJB has been supported in part by the Center for Advanced Studies at the University of Illinois.

\label{sec:acks}

%
\label{sec:bib}
\bibliographystyle{plain}
\bibliography{info490}

\begin{thebibliography}{10}

\bibitem{anscombe1973graphs}
F.~J. Anscombe.
\newblock Graphs in statistical analysis.
\newblock {\em The American Statistician}, 27(1):17--21, 1973.

\bibitem{asa}
American~Statistical Association.
\newblock \url{http://stat-computing.org/dataexpo/2009}.

\bibitem{brunner2013practical}
R.~J. Brunner.
\newblock Practical informatics: Training the next generation.
\newblock In {\em Astronomical Data Analysis Software and Systems XXII}, volume
  475, page~3, 2013.

\bibitem{cobb2015mere}
G.~W. Cobb.
\newblock Mere renovation is too little too late: We need to rethink our
  undergraduate curriculum from the ground up.
\newblock {\em arXiv preprint arXiv:1507.05346}, 2015.

\bibitem{davenport2012data}
T.~H. Davenport and D.~J. Patil.
\newblock Data scientist: the sexiest job of the 21st century.
\newblock {\em Harvard business review}, 90:70--76, 2012.

\bibitem{davidson-pilon2015bayesian}
C.~Davidson-Pilon.
\newblock {\em Bayesian Methods for Hackers: Probabilistic Programming and
  Bayesian Inference}.
\newblock Addison-Wesley Professional, 1st edition, 2015.

\bibitem{docker}
{Docker, Inc.}
\newblock \url{https://www.docker.com}.

\bibitem{docker-machine}
{Docker, Inc.}
\newblock \url{https://docs.docker.com/machine/}.

\bibitem{donoho2015fifty}
D.~Donoho.
\newblock 50 years of data science.
\newblock
  \url{http://courses.csail.mit.edu/18.337/2015/docs/50YearsDataScience.pdf}.

\bibitem{downey2012think}
A.~Downey.
\newblock {\em Think Python}.
\newblock {O'Reilly Media, Inc.}, 2012.

\bibitem{downey2013think}
A.~Downey.
\newblock {\em Think Bayes}.
\newblock {O'Reilly Media, Inc.}, 2013.

\bibitem{downey2014think}
A.~Downey.
\newblock {\em Think stats}.
\newblock {O'Reilly Media, Inc.}, 2014.

\bibitem{hardin2014data}
J.~Hardin, R.~Hoerl, N.~J. Horton, and D.~Nolan.
\newblock Data science in the statistics curricula: Preparing students to think
  with data.
\newblock {\em arXiv preprint arXiv:1410.3127}, 2014.

\bibitem{hunter2007}
J.~D. Hunter.
\newblock Matplotlib: A 2d graphics environment.
\newblock {\em Computing In Science \& Engineering}, 9(3):90--95, 2007.

\bibitem{mckinsey}
McKinsey~Global Institute.
\newblock
  \url{http://www.mckinsey.com/insights/business_technology/big_data_the_next_frontier_for_innovation}.

\bibitem{jupyter}
Project Jupyter.
\newblock \url{http://jupyter.org}.

\bibitem{nolan2012computing}
D.~Nolan and D.~Temple~Lang.
\newblock Computing in the statistics curricula.
\newblock {\em The American Statistician}, 2012.

\bibitem{patil2010pymc}
A.~Patil, D.~Huard, and C.~J. Fonnesbeck.
\newblock {PyMC}: Bayesian stochastic modelling in python.
\newblock {\em Journal of statistical software}, 35(4):1, 2010.

\bibitem{scikit-learn}
F.~Pedregosa, G.~Varoquaux, A.~Gramfort, V.~Michel, B.~Thirion, O.~Grisel,
  M.~Blondel, P.~Prettenhofer, R.~Weiss, V.~Dubourg, J.~Vanderplas, A.~Passos,
  D.~Cournapeau, M.~Brucher, M.~Perrot, and E.~Duchesnay.
\newblock Scikit-learn: Machine learning in {P}ython.
\newblock {\em Journal of Machine Learning Research}, 12:2825--2830, 2011.

\bibitem{boot2docker}
The~Boot2docker Project.
\newblock \url{http://boot2docker.io/}.

\bibitem{perez2007ipython}
F.~Pérez and B.~Granger.
\newblock Ipython: A system for interactive scientific computing.
\newblock {\em Computing in Science \& Engineering}, 9(3):21--29, 2007.

\bibitem{shotts2012linux}
W.~E. Shotts~Jr.
\newblock {\em The linux command line: A complete introduction}.
\newblock No Starch Press, 2012.

\bibitem{walt2011numpy}
S.~van~der Walt, S.~C. Colbert, and G.~Varoquaux.
\newblock The numpy array: A structure for efficient numerical computation.
\newblock {\em Computing in Science \& Engineering}, 13(2):22--30, 2011.

\bibitem{seaborn}
M.~Waskom.
\newblock \url{http://stanford.edu/~mwaskom/software/seaborn/}.

\bibitem{wickham2011asa}
H.~Wickham.
\newblock {ASA} 2009 data expo.
\newblock {\em Journal of Computational and Graphical Statistics}, 20(2), 2011.

\bibitem{yau}
N.~Yau.
\newblock
  \url{http://flowingdata.com/2009/11/12/how-to-make-a-us-county-thematic-map-using-free-tools}.

\end{thebibliography}

\end{document}